\documentclass{article}
\usepackage{spconf}
\usepackage{amsmath}
\usepackage{amsfonts}
\usepackage{amssymb}
\usepackage{scalerel}
\usepackage[hidelinks]{hyperref}
\usepackage[capitalize,nameinlink]{cleveref}
\usepackage{graphicx}
\usepackage{xspace}
\usepackage{multicol}
\usepackage[inline]{enumitem}
\usepackage{booktabs}
\usepackage{float}
\usepackage{siunitx}
\usepackage{tikz}
\usepackage{bm}
\usepackage{url}
\usepackage{pifont}
\usepackage{microtype}

\usetikzlibrary{shapes,arrows,positioning,perspective,fit,3d,backgrounds,decorations.pathreplacing,calc}

\newcommand{\pwconv}{PW\raisebox{1pt}{$\circledast$}\xspace}
\newcommand{\dwconv}{DDW\raisebox{1pt}{$\circledast$}}
\newcommand{\cmark}{\ding{51}}%
\newcommand{\xmark}{\ding{55}}%

\newcommand\blfootnote[1]{%
  \begingroup
  \renewcommand\thefootnote{}\footnote{#1}%
  \addtocounter{footnote}{-1}%
  \endgroup
}

\title{Scalable Speech Enhancement with Dynamic Channel Pruning}
\name{Riccardo Miccini$^{\star \dagger}$ \qquad Clément Laroche$^{\star}$ \qquad Tobias Piechowiak$^{\star}$ \qquad Luca Pezzarossa$^{\dagger}$}

\address{$^{\star}$ GN Audio \quad $^{\dagger}$ Technical University of Denmark}

\begin{document}
\ninept
\maketitle
\begin{abstract}
Speech Enhancement (SE) is essential for improving productivity in remote collaborative environments.
Although deep learning models are highly effective at SE, their computational demands make them impractical for embedded systems.
Furthermore, acoustic conditions can change significantly in terms of difficulty, whereas neural networks are usually static with regard to the amount of computation performed.
To this end, we introduce Dynamic Channel Pruning to the audio domain for the first time and apply it to a custom convolutional architecture for SE.
Our approach works by identifying unnecessary convolutional channels at runtime and saving computational resources by not computing the activations for these channels and retrieving their filters.
When trained to only use \qty{25}{\percent} of channels, we save \qty{29.6}{\percent} of MACs while only causing a \qty{0.75}{\percent} drop in PESQ.
Thus, DynCP offers a promising path toward deploying larger and more powerful SE solutions on resource-constrained devices.
\end{abstract}

\blfootnote{This work has received funding from the European Union’s Horizon research and innovation program under grant agreement No 101070374.}

\vspace{-2em}

\begin{figure*}[t]
\centering
\includegraphics[width=1.0\textwidth]{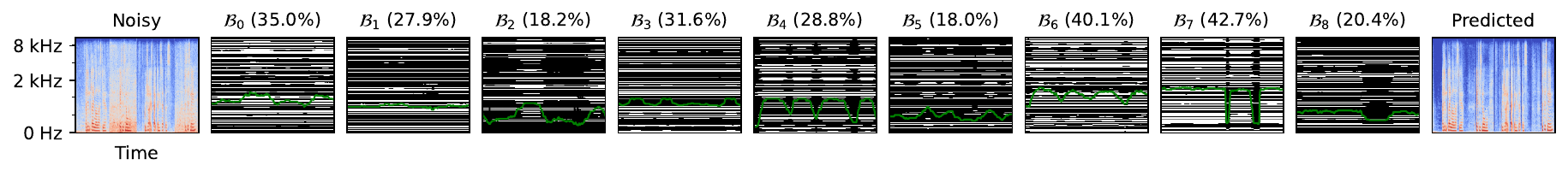}
\vspace{-2em}
\caption{Noisy input, channel states, and predicted speech computed from an \qty{8}{\second} sample; x-axis shows time, y-axis shows frequency for spectrograms and channel index for $\mathcal{B}_i$; white channels are kept, black channels are omitted; green lines and titles indicate the instantaneous and average pruning ratio for the given block, respectively.}
\label{fig:dyncp_sample}
\end{figure*}

\section{Introduction}
\label{sec:intro}
Real-time speech enhancement (SE) and noise suppression can facilitate communication in noisy environments and have become ubiquitous in devices such as speakerphones, earbuds, and headsets.
Thanks to the recent advances in deep learning, SE solutions based on neural networks made considerable strides, surpassing digital signal processing techniques in effectiveness~\cite{defossez_real_2020,jia_tfcn_2022}.
However, the constraints of the aforementioned embedded devices (in terms of energy, memory, and computation) are often too stringent to support state-of-the-art SE solutions based on deep learning. 
In particular, when deploying SE models on embedded devices, engineers and practitioners must not only comply with the constraints of the target platform but also try to maximize battery life, which further limits the computational complexity of the model, further compromising its effectiveness.
An undesired outcome of this design-time trade-off is that the deployed model may be inadequate or redundant, depending on the current acoustic conditions, which vary significantly in the real-world.

In this paper, we seek to postpone the trade-off between effectiveness and computational efficiency to inference time and outsource it to the model itself.
To this end, we train our model so that it can modulate the amount of computation performed based on the current input data.
This is similar to how humans use more cognitive resources to perform more difficult tasks and is an active area of research within the realm of Dynamic Neural Networks (DynNNs)~\cite{han_dynamic_2021}.
Under the right circumstances, a DynNN may skip parts of its computational graph and incur substantial savings in terms of computational resources such as memory transfers or arithmetic operations.

Previous works in the audio domain explored dynamism along the model \textit{depth}, represented by the number of computational blocks in the graph~\cite{chen_dont_2020,miccini_dynamic_2023,bralios_latent_2022}, as well as \textit{width} --- i.e. the number of convolutional channels in the intermediate activations~\cite{elminshawi_slim-tasnet_2023}.
In our work, we take a more fine-grained approach to width-based dynamism by allowing the model to select or exclude individual convolutional filters, thereby saving the computational costs associated with retrieving the weights of the omitted filters and computing their respective intermediate activations.
This idea, originally introduced in \cite{lin_runtime_2017}, has been extended in several works~\cite{hua_channel_2019,gao_dynamic_2019,rao_runtime_2019,spasov_dynamic_2019}, and is known as Dynamic Channel Pruning, Channel Gating, Dynamic Filter Selection, and more; throughout this text, we employ Dynamic Channel Pruning (DynCP) to indicate this spectrum of techniques.

Most notably, however, the majority of works on DynNNs and DynCP are constrained within the field of computer vision, meaning that the effectiveness of channel pruning on SE is yet to be leveraged or assessed.
Therefore, motivated by the potential benefits of DynNNs for resource-constrained speech enhancement, we present the following contributions:
\begin{enumerate*}[label={\arabic*)}]
    \item Introduce a fully-convolutional architecture based on depthwise-separable dilated convolution;
    \item Integrate a lightweight gating module that is trained jointly with the backbone to determine which channels can be skipped;
    \item Evaluate the dynamic architecture on a popular dataset for speech enhancement and noise suppression;
    \item Analyze and discuss the impact of different hyperparameters and training strategies on model performance.
\end{enumerate*}

\section{Related Work}
\label{sec:related}
Dynamic Neural Networks are a class of neural networks where some aspects of the computational graph change at inference-time, usually based on the input data.
We refer the interested reader to \cite{han_dynamic_2021} for a comprehensive survey of DynNN techniques.
In this section, we focus exclusively on sample-wise techniques applied to audio applications where the depth or width of the graph is modulated.
Furthermore, we broaden the meaning of DynNNs to include works where the changes in the graph are controlled by the user.

\begin{description}[
    font=\bfseries,
    leftmargin=0pt,
    parsep=\parsep,
    listparindent=\parindent,
    labelwidth=0em,
    itemindent=1em,
    labelsep=1em,
    align=left,
    itemsep=\parsep,
]
\item[Depth] 
We only employ a subset of the layers in the model. 
With \textit{Early-Exiting}, we accept the output of a given intermediate layer as our result and do not execute the remaining layers; in \cite{miccini_dynamic_2023}, an established architecture for noise suppression has been adapted to perform manual early-exiting by extending its reconstruction loss to the intermediate activations while providing a secondary path for richer internal representations.
Conversely, \cite{chen_dont_2020,li_learning_2021} fully automate the process by exiting when the output of the given layer is too similar to the output of the previous one.
In \textit{Recursive} networks, we imitate a deeper model by executing each layer more than once; in \cite{bralios_latent_2022}, a speech separation backbone is trained end-to-end alongside a gating module to determine the ideal number of iterations per layer.

\item[Width] 
Only a subset of nodes in each layer are used or computed. 
In the context of 1D or 2D convolutional layers, these correspond to specific channels of their output tensor.
Therefore, by excluding a given convolutional channel from the graph, we save the costs associated with retrieving its weights and computing its activations. 
\textit{Slimmable} networks are trained to support several ``utilization factors'' --- i.e. different numbers of active channels per layer --- resulting is an ensemble of weight-sharing subnets with different performance/efficiency trade-offs; \cite{elminshawi_slim-tasnet_2023} adapts this mechanism to the popular source separation architecture Conv-TasNet~\cite{luo_conv-tasnet_2019} allowing the user to manually select the desired subnet; subsequently, in~\cite{elminshawi_dynamic_2024}, an end-to-end extension with adaptive utilization factor is applied to a state-of-the-art speech separation network.
\textit{Dynamic Channel Pruning} architectures take a more fine-grained approach, allowing each convolutional channel to be selected or omitted independently according to a binary mask.
The latter can be computed adaptively, e.g. by solving a resource-allocation problem~\cite{spasov_dynamic_2019} or through a gating subnet~\cite{hua_channel_2019}. 
To the best of our knowledge, this technique has not been applied to audio-to-audio processing tasks prior to this work.
\end{description}


\section{The Conv-FSENet Architecture}
\label{sec:convfsenet}

\subsection{Problem formulation}
\label{ssec:problem}
We consider the problem of single-channel speech enhancement in the time-frequency domain, where our input signal $x(t)$ is a mixture of target speech $s(t)$ and background noise $n(t)$.
In the Short-Time Fourier Transform (STFT) domain, we have:
\begin{equation}
X(l,f) = S(l,f) + N(l,f)
\end{equation}
where $l$ is the STFT frame index and $f$ is the frequency index.
We estimate the target speech spectrum $\widehat{S}$ by applying a mask $\widehat{M}$ to the complex spectrogram of our noisy input:
\begin{equation}
\widehat{S}(l,f) = X(l,f) \cdot \widehat{M}(l,f)
\end{equation}
Although time-domain architectures are proving effective, we rely on spectrogram data to simplify integration within audio pipelines.

\subsection{Model Architecture}
\label{ssec:arch}

\begin{figure*}[t]
\centering
\resizebox{0.98\textwidth}{!}{\begin{tikzpicture}[
    node distance=1em,
    auto,
    scale=0.75,
]

\tikzset{every node/.append style={transform shape}}
\tikzset{box/.style={rectangle, draw, semithick, align=center, text centered, inner sep=0.5em, fill=red!35, minimum height=2.5em}}
\tikzset{boxdots/.style={box, fill=none, draw=none, inner sep=0em}}
\tikzset{group/.style={rectangle, rounded corners, draw, semithick, text centered, inner sep=0.5em, fill=blue!10}}
\tikzset{tcngroup/.style={rectangle, rounded corners, fill=none, thick, text centered, densely dashed, inner sep=1em, draw=red!35}}
\tikzset{proc/.style={rectangle, draw, rounded corners, semithick, text centered, inner sep=0.5em, fill=white, minimum height=2em}}
\tikzset{inout/.style={minimum width=1em}}
\tikzset{ar/.style={draw, semithick, ->}}

\node [inout] (in) {$x$};
\node [proc, right=of in] (stft) {STFT};
\node [proc, right=2em of stft] (mag) {$\left|\cdot\right|$};
\node [box, right=1.75em of mag, fill=olive!25] (fec) {\pwconv};
\node [box, right=of fec, fill=yellow!25] (fea) {ReLU};
\node [box, right=3em of fea] (bl1) {$\mathcal{B}_1$};
\node [boxdots, right=0.5em of bl1] (bl2) {$\boldsymbol{\cdots}$};
\node [box, right=0.5em of bl2] (bl3) {$\mathcal{B}_{N_b}$};
\node [box, right=of bl3, fill=yellow!25] (blr) {ReLU};

\node [boxdots, right=1.5em of blr] (bl4) {$\boldsymbol{\cdots}$};

\node [boxdots, right=1.5em of bl4] (bl5) {$\boldsymbol{\cdots}$};
\node [box, right=0.5em of bl5] (bl6) {$\mathcal{B}_{N_b N_s}$};

\node [box, right=3em of bl6, fill=olive!25] (bec) {\pwconv};
\node [box, right=of bec, fill=yellow!25] (bea) {Sigmoid};
\node [proc, circle, inner sep=0em, minimum height=1.5em, below right=3em and 1.25em of bea.east] (mult) {$\times$};
\node [proc, right=2em of mult] (istft) {iSTFT};
\node [inout, right=of istft] (out) {$\hat{s}$};

\draw [decorate,decoration={brace,amplitude=5pt,raise=0.2em}]
  (bl1.north west) -- (bl3.north east) node[midway,yshift=1.25em,inner sep=0em] (nb) {$N_b$};

\draw [decorate,decoration={brace,amplitude=5pt,raise=0.25em}]
  (bl5.north west) -- (bl6.north east) node[midway,yshift=1.25em,inner sep=0em] (nbd) {$N_b$};

\node[above left=0.5em and 0em of fea.north east,inner sep=0em] (fel)  {\textbf{Front-End}};
\node[above left=0.5em and 0em of bea.north east,inner sep=0em] (bel) {\textbf{Back-End}};

\begin{scope}[on background layer]
\node [group, fit={(fec) (fea) (fel)}] (fe) {};
\node [group, fill=red!15, fit={(bl1) (blr) (nb)}] (s1) {};
\node [group, fill=red!15, fit={(bl5) (bl6) (nbd)}] (s3) {};
\node [group, fit={(bec) (bea) (bel)}] (be) {};
\end{scope}

\node[below left=0em of s1.north east,inner sep=0.5em]  {\textbf{Stack}};
\draw [decorate,decoration={brace,amplitude=5pt,raise=0.25em}]
  (s1.north west) -- (s3.north east) node[midway,yshift=1.25em,inner sep=0em] (ns) {$N_s$};

\begin{scope}[on background layer]
\node [tcngroup, fit={(bl1) (bl6) (ns.south)}] (tcn) {};
\end{scope}
\node[below left=0em of tcn.north east,inner sep=0.5em]  {\textbf{TCN}};

\draw [ar] (in) -- (stft);
\draw [ar] (stft) -- (mag) node[midway,yshift=0.3em,inner sep=0em] {$X$};
\draw [ar,shorten >=0.5em] (mag) -- (fec);
\draw [ar] (fec) -- (fea);

\draw [ar, shorten >=0.5em, shorten <=0.5em] (fea) -- (bl1);
\draw [ar, shorten >=0.5em, shorten <=0.5em] (bl6) -- (bec);

\draw [ar] (bl3) -- (blr);

\draw [ar] (bec) -- (bea);
\draw [ar, shorten <=0.5em] (bea.east) -| (mult.north) node[midway,xshift=0.5em,inner sep=0em] {$\widehat{M}$};
\draw [ar] (stft.east) -| ++(1em, 0em) |- (mult.west);
\draw [ar] (mult) -- (istft) node[midway,yshift=0.3em,inner sep=0em] {$\widehat{S}$};
\draw [ar] (istft) -- (out);

\end{tikzpicture}}
\vspace{-0.75em}
\caption{Overall architecture of Conv-FSENet.}
\label{fig:convfsenet}
\end{figure*}
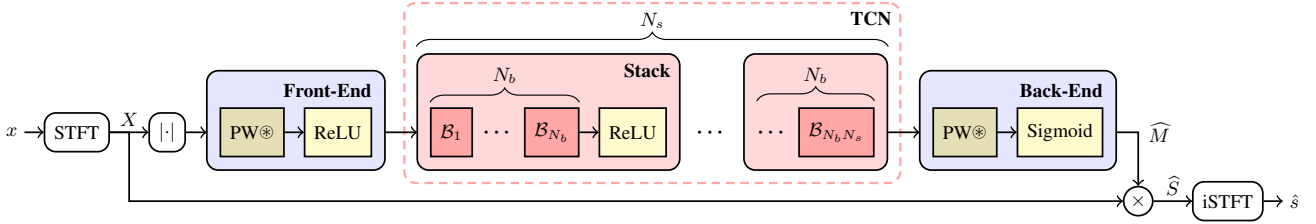

\begin{table}[t]
\vspace{-6pt}
\centering
\caption{Overview of the notation.}
\label{tab:params}
\vspace{2pt}
\resizebox{1.0\columnwidth}{!}{
\begin{tabular}{ll}
\toprule
\textbf{Notation} & \textbf{Description} \\
\midrule
$L$, $L_\text{RF}$ \hspace{1em} & Lengths of training input sequences and receptive field \\
$N_b$ & Number of blocks per stack \\
$N_s$ & Number of stacks \\
$C_\text{res}$ & Convolutional channels in residual branch \\
$C_\text{conv}$ & Convolutional channels inside blocks \\
$C_\text{gate}$ & Hidden features in gating module \\
$k$ & Kernel size for depthwise convolution \\
\midrule
$\mathcal{B}_i$ & Depthwise-separable convolutional block \\
$\mathcal{G}_i$ & Gating module for dynamic channel pruning \\
\pwconv & Point-wise convolution ($k=1$) \\
\dwconv & Dilated depth-wise convolution (\num{1} filter per channel)\\
$\text{P}(\cdot)$ & Time-pooling function \\
$\text{H}(\cdot)$ & Binarization function (variants of Heaviside) \\
$\Phi_\text{trgt}$ & Target pruning ratio (fraction of active channels) \\
\bottomrule
\end{tabular}
}
\end{table}

The architecture presented here is inspired by similar works on source separation and speech enhancement~\cite{luo_conv-tasnet_2019,pandey_tcnn_2019}.
To hint at the similarities and differences with Conv-TasNet~\cite{luo_conv-tasnet_2019}, we named it \textit{Convolutional Frequency-Domain Speech Enhancement Network} (Conv-FSENet).
As shown in \cref{fig:convfsenet}, it comprises a front-end, a sequence-modeling network based on Temporal Convolutional Networks (TCN)~\cite{bai_empirical_2018}, and a back-end.
Its parameters and notation are detailed in \cref{tab:params}; any convolution mentioned here refers to 1D.

Compared to traditional sequence models such as Gated Recurrent Units (GRU) or Long Short-Term Memory (LSTM), TCNs can process the entire input sequence simultaneously because new outputs do not depend on previous ones.
Additionally, its modular design and configurability let us meet a wide range of computational budgets, consequently resulting in different performances on the task.
Specifically, increasing the depth (number of blocks and stacks) widens the context window --- also called \textit{receptive field} or $L_\text{RF}$ --- of the model thereby enhancing its ability to capture long-range dependencies and complex patterns, while increasing the width (number of convolutional channels) promotes richer internal feature representations. 

In Conv-FSENet, a front-end takes the magnitude of the input STFT and applies \pwconv along with ReLU to map $F$ frequency bins into $C_\text{res}$ channels.
The TCN consists of a sequence of blocks $\mathcal{B}_i$ arranged into $N_s$ stacks of $N_b$ blocks each.
Within a stack, we double the dilation rate of each consecutive block, starting from \num{1}, and up to $2^{N_b}$.
Each stack except the last ends with a ReLU; the stacks are repeated sequentially $N_s$ times.
Subsequently, a back-end derives the denoising mask $\widehat{M}$ using a \pwconv and sigmoid activation.
Finally, the predicted clean speech spectrum $\widehat{S}$ is obtained by multiplying the mask with the complex-valued STFT of the input.

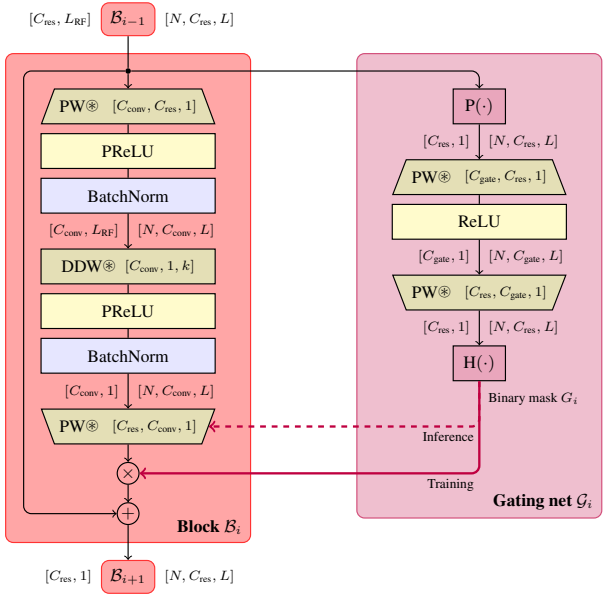
\begin{figure}[t]
\centering
\resizebox{0.93\columnwidth}{!}{\begin{tikzpicture}[
    node distance=1em,
    auto,
]

\tikzset{every node/.append style={transform shape}}
\tikzset{box/.style={rectangle, draw, semithick, align=center, text centered, inner sep=0.25em, minimum height=2em}}
\tikzset{cbox/.style={box, fill=olive!25, minimum width=10em, trapezium stretches=true}}
\tikzset{abox/.style={cbox, fill=yellow!25, minimum width=10em, minimum height=2em}}
\tikzset{bbox/.style={abox, fill=blue!10}}
\tikzset{gbox/.style={box, fill=purple!35, minimum width=3em}}
\tikzset{ann/.style={node distance=0.25em and 0em, inner sep=0em, minimum height=1.1em, fill=none}}
\tikzset{group/.style={rectangle, rounded corners, draw=red!80, semithick, inner xsep=2em, inner ysep=1em, fill=red!35}}
\tikzset{ggroup/.style={group, inner xsep=2em, inner ysep=2em, fill=purple!25, draw=purple!70}}
\tikzset{inout/.style={group, inner sep=0.5em}}

\tikzset{arr/.style={draw, semithick, rounded corners}}
\tikzset{ar/.style={arr, ->}}
\tikzset{arg/.style={ar, draw=purple, very thick}}

\node [circle,fill,inner sep=1pt] (inter) {};
\node [inout, above=2em of inter.center] (in) {$\mathcal{B}_{i-1}$};
\node [inner sep=0.5em, right=0em of in] (an1) {\scriptsize $[N, C_\text{res}, L]$};
\node [inner sep=0.5em, left=0em of in] (an1) {\scriptsize $[C_\text{res}, L_\text{RF}]$};

\node [cbox, shape=trapezium, trapezium angle=60, below=of inter.center] (pw1) {\pwconv\enspace\raisebox{.25\height}{\scriptsize $[C_\text{conv}, C_\text{res}, 1]$}};
\node [abox, below=0.5em of pw1] (act1) {PReLU};
\node [bbox, below=0.5em of act1] (bn1) {BatchNorm};

\node [cbox, below=2em of bn1] (dw) {\dwconv\enspace\raisebox{.25\height}{\scriptsize $[C_\text{conv}, 1, k]$}};
\node [abox, below=0.5em of dw] (act2) {PReLU};
\node [bbox, below=0.5em of act2] (bn2) {BatchNorm};

\node [cbox, shape=trapezium, trapezium angle=-60, below=2em of bn2] (pw2) {\pwconv\enspace\raisebox{.25\height}{\scriptsize $[C_\text{res}, C_\text{conv}, 1]$}};

\node [box, below=of pw2, circle, inner sep=0em, minimum height=1.25em] (prod) {$\times$};
\node [box, below=of prod, circle, inner sep=0em, minimum height=1.25em] (sum) {$+$};

\node [inout, below=2em of sum] (out) {$\mathcal{B}_{i+1}$};
\node [inner sep=0.5em, right=0em of out] (an2) {\scriptsize $[N, C_\text{res}, L]$};
\node [inner sep=0.5em, left=0em of out] (an2) {\scriptsize $[C_\text{res}, 1]$};

\node [gbox, right=14em of pw1] (pool) {$\text{P}(\cdot)$};
\node [cbox, shape=trapezium, trapezium angle=60, below=2em of pool] (gpw1) {\pwconv\enspace\raisebox{.25\height}{\scriptsize $[C_\text{gate}, C_\text{res}, 1]$}};
\node [abox, below=0.5em of gpw1] (gact1) {ReLU};
\node [cbox, shape=trapezium, trapezium angle=-60, below=2em of gact1] (gpw2) {\pwconv\enspace\raisebox{.25\height}{\scriptsize $[C_\text{res}, C_\text{gate}, 1]$}};
\node [gbox, below=2em of gpw2] (bin) {$\text{H}(\cdot)$};
\node [circle,fill=none,inner sep=1pt, below=5.5em of bin] (inter2) {};

\begin{scope}[on background layer]
\node [group, fit={(inter.center) (dw) (sum)}] (b) {};
\node [ggroup, fit={(inter2) (pool) (gact1) (bin)}] (g) {};
\end{scope}

\node[above left=0em of b.south east,inner sep=0.5em]  {\textbf{Block $\mathcal{B}_i$}};
\node[above left=0em of g.south east,inner sep=0.5em]  {\textbf{Gating net $\mathcal{G}_i$}};

\draw [arr] (in) -- (inter);
\draw [ar] (inter) -- (pw1);
\draw [arr] (pw1) -- (act1);
\draw [arr] (act1) -- (bn1);

\draw [ar] (bn1) -- (dw) node[midway,right,inner sep=0.5em] {\scriptsize $[N, C_\text{conv}, L]$} node[midway,left,inner sep=0.5em] {\scriptsize $[C_\text{conv}, L_\text{RF}]$};
\draw [arr] (dw) -- (act2);
\draw [arr] (act2) -- (bn2);

\draw [ar] (bn2) -- (pw2) node[midway,right,inner sep=0.5em] {\scriptsize $[N, C_\text{conv}, L]$} node[midway,left,inner sep=0.5em] {\scriptsize $[C_\text{conv}, 1]$};

\draw [ar] (pw2) -- (prod);
\draw [ar] (prod) -- (sum);
\draw [ar] (sum) -- (out);

\draw [ar] (inter) -| ++(-6em,-1em) |- (sum.west);

\draw [ar] (inter) -| (pool.north);
\draw [ar] (pool) -- (gpw1) node[midway,right,inner sep=0.5em] {\scriptsize $[N, C_\text{res}, L]$} node[midway,left,inner sep=0.5em] {\scriptsize $[C_\text{res}, 1]$};
\draw [arr] (gpw1) -- (gact1);
\draw [ar] (gact1) -- (gpw2) node[midway,right,inner sep=0.5em] {\scriptsize $[N, C_\text{gate}, L]$} node[midway,left,inner sep=0.5em] {\scriptsize $[C_\text{gate}, 1]$};
\draw [ar] (gpw2) -- (bin) node[midway,right,inner sep=0.5em] {\scriptsize $[N, C_\text{res}, L]$} node[midway,left,inner sep=0.5em] {\scriptsize $[C_\text{res}, 1]$};
\draw [arg,dashed] (bin) |- (pw2) node[midway,below,inner sep=0.25em,anchor=north east] {\scriptsize Inference};
\draw [arg] (bin) |- (prod) node[midway,below,inner sep=0.25em,anchor=north east] {\scriptsize Training};
\draw [ar,draw=none] (bin.south) -- ++(0em,-2em) node[midway,right,inner sep=0.5em] {\scriptsize Binary mask $G_i$};



\end{tikzpicture}}
\caption{Structure of baseline block $\mathcal{B}_i$ (red box) and gating module $\mathcal{G}_i$ (purple box) for DynCP; annotations along the graph refer to activation shapes (right: training, left: streaming inference), while those inside the convolutional layers refer to weights.}
\label{fig:block_dyncp}
\end{figure}

The structure of the blocks $\mathcal{B}_i$ is illustrated in \cref{fig:block_dyncp} (red box).
Similarly to \cite{pandey_tcnn_2019}, each block performs residual dilated depthwise-separable convolution with interwoven non-linearity and normalization --- a breakdown of these attributed follows below.
Depthwise-separable convolution decomposes a regular convolution into \pwconv, equivalent to applying a dense layer to each time step independently, and \dwconv, where each channel is convolved by its own separate filter of size $k$.
This substantially decreases the number of parameters and operations.
We apply a dilation rate to each \dwconv to increase its receptive field without introducing additional blocks or trainable parameters.
We can make \dwconv causal --- i.e., not dependent on future time steps --- by padding its input and cropping its output accordingly. 
Lastly, we map our $C_\text{conv}$ internal channels back into $C_\text{res}$ channels with another \pwconv.
Here, a skip connection causes each block to learn a residual function with respect to its input to facilitate the training of deeper networks.

As in \cite{Braun_2020,miccini_dynamic_2023}, our loss function is the square of the difference between the reconstructed and original clean speech spectra, computed in both the absolute magnitude and complex domains, each weighted according to $\alpha$ and with their dynamic range compressed by $c$:
\begin{equation}
\label{eq:loss_se}
\mathcal{L}_\text{SE}^{} = \alpha \! \sum_{l, f} \left| |S|^c e^{j \theta_{\scaleto{\!S\rule{0ex}{1.5ex}}{0.5em}}} \! - |\widehat{S}|^c e^{j \theta_{\scaleto{\!\widehat{S}\rule{0ex}{1.5ex}}{0.5em}}} \right|^{\scriptscriptstyle 2} \! + (1-\alpha) \! \sum_{l, f} \left| |S|^c \! - |\widehat{S}|^c\right|^{\scriptscriptstyle 2}
\end{equation}

\section{Dynamic Channel Pruning}
\label{sec:method}
The number of multiply-accumulate operations (MACs) performed by each type of convolution can be approximated as:
\begin{equation}
\text{MAC}_\text{\,\pwconv} \approx L \cdot C_\text{conv} \cdot C_\text{res}
\qquad
\text{MAC}_\text{\,\dwconv} \approx L \cdot C_\text{conv} \cdot k
\end{equation}
Since $C_\text{res} \gg k$, we will ignore the impact of the middle \dwconv and only concentrate on the last \pwconv, leaving the other layers unaffected.
To determine which convolutional channels to compute, we pair each block $\mathcal{B}_i$ with a gating module $\mathcal{G}_i$, shown in purple in \cref{fig:block_dyncp}.
This subnet generates a binary gating mask $G_i \in \left\{0, 1\right\}^{C_\text{res}}$ from the block input, where \num{0} means that a given channel is omitted and \num{1} means it is kept.
Due to the skip connections, each omitted channel will maintain the value computed by the previous block.

During training, the computational graph remains static and we multiply the mask $G_i$ with the output of the last \pwconv in the block before adding the skip connection, as indicated by the continuous purple line in \cref{fig:block_dyncp}.
During inference, however, we use the mask to decide which channels in the last \pwconv layer (connected to the dotted purple line) are active.
In this case, the activations of the omitted channels are not computed and their corresponding filters are not retrieved, resulting in computational savings at runtime.
The gating module $\mathcal{G}_i$ comprises the following three parts: 

\begin{description}[
    font=\bfseries,
    leftmargin=0pt,
    parsep=\parsep,
    listparindent=\parindent,
    labelwidth=0em,
    itemindent=1em,
    labelsep=1em,
    align=left,
    itemsep=\parsep,
]
\item[Pooling function $\text{P}(\cdot)$]
We aggregate information from multiple time steps by computing the moving average of our intermediate activations over $L_\text{pool}$ frames.
During real-time streaming, this can be efficiently approximated with a first-order IIR filter parameterized by a smoothing coefficient $\beta$ such that $\text{P}(x_t) = \beta x_t + (1-\beta) \text{P}(x_{t-1})$.

\item[Processing stack]
This computes raw pruning scores from the sequence averages. It consists of two \pwconv (equivalent to dense during real-time streaming) and a ReLU activation.

\item[Binarization function $\text{H}(\cdot)$]
Finally, we obtain $\mathcal{G}_i$ from the raw scores by applying the Heaviside step function.
Since its derivative is \num{0} over the entire domain (except for $x=0$), we must devise a way to backpropagate through it during training.
To this end, we employ three strategies based on surrogate gradient (for Sigmoid and SuperSpike~\cite{nilsson_resource-efficient_2024}, respectively) and a stochastic approach based on the ``binary special case'' of the Concrete distribution~\cite{maddison_concrete_2017}.
\end{description}

Finally, we enforce the desired dynamic behavior through an auxiliary loss term, optimized jointly with \cref{eq:loss_se}.
Specifically, we want to minimize the difference between the training-time binary mask tensor $G \in \left\{0, 1\right\}^{N \times C_\text{res} \times L \times I}$ (where $I = N_s N_b$ is the total number of blocks) and a target pruning ratio $\Phi_\text{trgt}$.
To simplify the learning process, we average along the batch, time, and block dimensions before computing the mean squared error:
\begin{equation}
\label{eq:loss_dyncp}
\mathcal{L}_\text{DCP}^{} =\frac{1}{C_\text{res}} \sum_c^{C_\text{res}} \! \left( \frac{1}{N L I} \! \sum_{n}^{N} \sum_{l}^{L} \sum_{i}^{I} \! G_{n, c, l, i} - \Phi_\text{trgt} \right)^2
\end{equation}
This provides regularization, smoothing and reducing the possible directions of the gradient, while causing the model to prioritize global behavior instead of local consistency when minimizing the loss.

\section{Experimental setup}
\label{sec:exp_setup}

\begin{description}[
    font=\bfseries,
    leftmargin=0pt,
    parsep=\parsep,
    listparindent=\parindent,
    labelwidth=0em,
    itemindent=1em,
    labelsep=1em,
    align=left,
    itemsep=\parsep,
]

\item[Datasets] 
We trained and evaluated our models on the VoiceBank+DEMAND Dataset~\cite{valentini-botinhao_investigating_2016}, consisting of \num{11752} pairs of noisy-clean samples from 28 speakers with SNR between \qty{15}{\decibel} and \qty{0}{\decibel}.
Similarly, the test set includes \num{824} samples from two other speakers mixed with unseen noise at SNR between \qty{17.5}{\decibel} and \qty{2.5}{\decibel}.
All the data is downsampled to \qty{16}{\kilo\hertz}.
During training, the audio samples are split into random segments of \qty{4}{\second}; we employ the spectra augmentation and level invariance techniques described in \cite{Braun_2020}.
Finally, we apply an STFT with a window length of \num{512} samples and \qty{50}{\percent} overlap, resulting in $F=257$ features.

\item[Model] 
We tested our Conv-FSENet with the following parameters: $C_\text{res} = 128$, $C_\text{conv} = 256$, $k = 3$, $N_b = 3$, and $N_s = 3$ unless noted otherwise.
Additionally, in the DynCP variants, we have $C_\text{gate} = 16$, $L_\text{pool} = L_\text{RF}$ (i.e. the model receptive field), and $\Phi_\text{trgt} = 0.25$ unless noted otherwise.
In \cref{eq:loss_se} we use $\alpha = 0.3$ and $c = 0.3$.
\cref{fig:pareto,fig:dyncp_sample} are derived from non-causal specimens using the SuperSpike surrogate gradient~\cite{nilsson_resource-efficient_2024} to backpropagate through $\text{H}(\cdot)$. Causal variants and other binarization approaches are presented in \cref{tab:comparison}.

\item[Training] 
We train all our models using the Adam optimizer, with a learning rate of \num{1e-3} and weight decay of \num{1e-5}, on batches of \num{64} elements.
The static baselines are trained for \num{400} epochs.
We then fine-tune the dynamic networks starting from the pre-trained weights of each equivalently-sized baseline and random weights for $\mathcal{G}_i$, for an additional \num{120} epochs.
In both cases, we interrupt the training after \num{20} epochs without improvement and decay our learning rate by a factor of \num{0.5} after \num{3} validation rounds with no improvement.
We compute our validation metrics every \num{2} epochs.

\item[Metrics] 
We evaluate our solutions using the Perceptual Evaluation of Speech Quality (PESQ)~\cite{rix_perceptual_2001}, the Scale-Invariant Signal-to-Distortion Ratio (SI-SDR)~\cite{roux_sdr_2018}, and the MACs\footnote{Computed using \url{https://github.com/facebookresearch/fvcore/blob/main/docs/flop_count.md}} per STFT frame.
\end{description}

\section{Results}
\label{sec:results}

\begin{figure}[t]
\centering
\includegraphics[width=1.0\columnwidth]{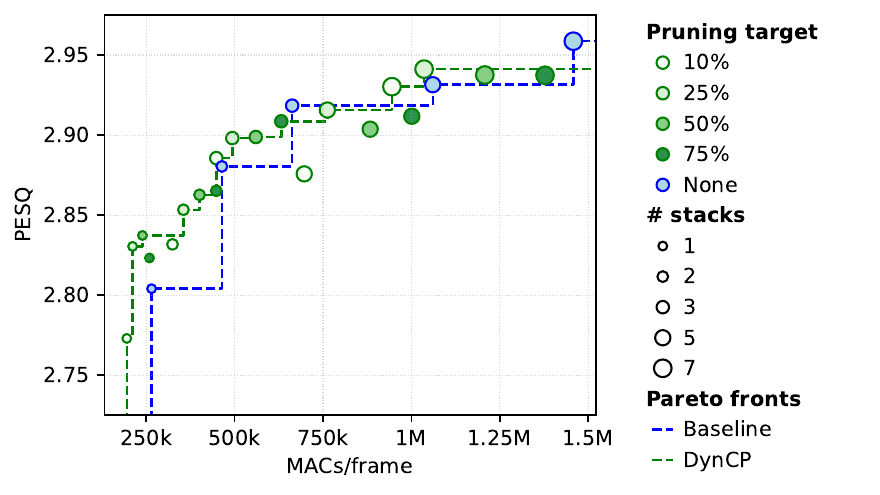}
\vspace{-2em}
\caption{Pareto fronts of PESQ vs. MACs for static baselines and DynCP models over a range of $N_s$ and $\Phi_\text{trgt}$ values.}
\label{fig:pareto}
\end{figure}

\begin{table}[t]
\vspace{-6pt}
\centering
\caption{Comparison of different backpropagation strategies.}
\label{tab:comparison}
\vspace{2pt}
\resizebox{1.0\columnwidth}{!}{
\begin{tabular}{lc|cc|cc}
\toprule
\textbf{Model} & \textbf{Causal} & \textbf{PESQ} & \textbf{SI-SDR} & \textbf{kMACs} & \textbf{Reduction}\\
\midrule
Noisy &  --- & \num{1.98} & \qty{8.45}{\decibel} & --- & --- \\
\midrule
Baseline & \xmark & \num{2.92} & \qty{17.92}{\decibel} & \num{662.78} & --- \\
DynCP (Sigmoid surrogate) & \xmark & \num{2.89} & \qty{17.82}{\decibel} & \num{484.81} & \qty{30.82}{\percent} \\
DynCP (SuperSpike surrogate) & \xmark & \num{2.90} & \qty{18.17}{\decibel} & \num{493.36} & \qty{29.60}{\percent} \\
DynCP (Binary Concrete distr.) & \xmark & \num{2.58} & \qty{18.32}{\decibel} & \num{457.75} & \qty{34.68}{\percent} \\
\midrule
Baseline & \cmark & \num{2.77} & \qty{17.54}{\decibel} & \num{663.02} &       --- \\
DynCP (Sigmoid surrogate) & \cmark & \num{2.73} & \qty{17.50}{\decibel} & \num{483.53} & \qty{31.03}{\percent} \\
DynCP (SuperSpike surrogate) & \cmark & \num{2.73} & \qty{17.52}{\decibel} & \num{485.05} & \qty{30.81}{\percent} \\
DynCP (Binary Concrete distr.) & \cmark & \num{2.68} & \qty{17.80}{\decibel} & \num{475.41} & \qty{32.18}{\percent} \\
\bottomrule
\end{tabular}
}
\end{table}

In \cref{fig:pareto}, we relate the denoising performances and computational efficiency of the Conv-FSENet static baselines with their DynCP counterparts.
Although our dynamic variants experience a drop in PESQ, we benefit from a significant reduction in MACs (between \qty{7}{\percent} and \qty{39}{\percent}, depending on $\Phi_\text{trgt}$), which makes most of our DynCP models more Pareto-efficient.
This is particularly true when $N_s \le 3$.
We recall that, in our experiments, the gating modules $\mathcal{G}_i$ use a pooling window equal to the receptive field $L_\text{RF} = N_s (k - 1) \sum^{N_b}_{n=1} 2^{n-1} + 1$.
Since this former is directly proportional to the number of stacks, we posit that smaller networks react faster to changes in input, exhibiting higher adaptiveness.
On the other hand, deeper networks perform pooling over a longer interval: for instance, with $N_s = 7$ we have a window of \qty{\sim 1.6}{\second}, which may hinder dynamism.

We demonstrate the dynamic behavior of our models in \cref{fig:dyncp_sample}.
Here, we created a synthetic signal composed of \qty{2}{\second} fragments from \num{4} distinct test samples, each with a unique combination of speaker, noise type, and SNR.
This input signal, showcasing various acoustic conditions, helps us understand how the different convolutional channels contribute to the task and how they are selected.
In particular, the number of active channels in blocks $\mathcal{B}_0$, $\mathcal{B}_1$, $\mathcal{B}_2$, and $\mathcal{B}_8$ correlates with the presence of louder noise, i.e. the regions of the noisy spectrogram with more intense high-frequency content.
Conversely, channels in $\mathcal{B}_3$, $\mathcal{B}_4$, $\mathcal{B}_6$, and $\mathcal{B}_7$ exhibit a strong correlation with the presence of speech.
In all blocks, however, there appears to be a subset of channels that are almost always active and a subset of channels that are never used.
A possible interpretation of the latter behavior may be found in how the gradients propagate through the model.
In multiplicative nodes --- such as the one in \cref{fig:block_dyncp} where the binary mask $G_i$ is applied to the output of the last \pwconv during training --- the incoming gradient is passed onto each input branch, after being scaled by the value in the other input branch.
Thus, the gradient of currently inactive channels is reduced to \num{0}, meaning that their respective filters will not co-adapt with the rest of the network, whereas the $\mathcal{G}_i$ branch will continue to learn.

In \cref{tab:comparison}, we show the effect of the backpropagation strategies described in \cref{sec:method}.
A possible solution consists of turning $\text{H}(\cdot)$ into a stochastic process sampling from a discrete distribution and training it using a continuous and differentiable relaxation (called Binary Concrete distribution in \cite{maddison_concrete_2017}).
The noise in $G_i$ would allow neglected channels to occasionally receive gradient and learn useful transformations along with the rest of the model.
Nevertheless, the drop in performances between the surrogate gradients and the aforementioned stochastic approach suggests that further work is needed.
Additionally, \cref{tab:comparison} shows performances on causal models.
Here, we observe a \qty{\sim 5}{\percent} drop in PESQ from the non-causal counterparts, in line with similar work~\cite{defossez_real_2020,jia_tfcn_2022}, whereas the DynCP models maintain a similar amount of MAC reduction. 

Lastly, since the gating modules expose unnecessary filters during training, we can use this information to perform static pruning and further optimize the model for deployment, with storage savings approximately equivalent to the percentage of MACs reduction.
Furthermore, each $\mathcal{G}_i$ can be adjusted to only predict gating values for channels whose activity is prone to change.  
If applied to the model used in \cref{fig:block_dyncp}, this simple heuristic would cause the average reduction in MACs to go from \qty{29.6}{\percent} to \qty{31.3}{\percent}.

\section{Conclusion}
\label{sec:conclusion}
We presented ConvFSE-Net, a neural network architecture for SE, and extended it with a dynamic pruning system that learns to skip unnecessary convolutional channels based on the input data.
Compared to the static baseline in \cref{tab:comparison}, our dynamic models can save up to \qty{29.6}{\percent} of MACs while only incurring a \qty{0.75}{\percent} drop in PESQ.
Thus, our results indicate a path towards higher efficiency, thanks to the model's ability to scale its computation adaptively.
We aim to extend this dynamic approach with alternative learning strategies and efficiency constraints.
Furthermore, we will apply this mechanism to more complex networks to verify its general applicability.


\bibliographystyle{IEEEbib}
\bibliography{refs}

\end{document}